\documentclass[aps,prl,twocolumn,showpacs]{revtex4}
\input{epsf}
\usepackage{epsfig}
\begin{document}

\title{Entangled light pulses from single cold atoms}

\author{Giovanna Morigi,$^1$ J\"urgen Eschner,$^2$ Stefano Mancini,$^3$ and David Vitali$^3$}
\affiliation{$^1$Grup d'Optica, Departament de Fisica, Universitat
Autonoma de
Barcelona, 08193 Bellaterra, Spain\\
$^2$ICFO - Institut de Ci\`{e}ncies Fot\`{o}niques, 08860 Castelldefels (Barcelona), Spain\\
$^3$Dipartimento di Fisica, Universit\`a di Camerino, 62032
Camerino, Italy}

\date{\today}

\begin{abstract}
The coherent interaction between a laser-driven single trapped
atom and an optical high-finesse resonator allows to produce
entangled multi-photon light pulses on demand. The mechanism is
based on the mechanical effect of light. The degree of
entanglement can be controlled through the parameters of the laser
excitation. Experimental realization of the scheme is within reach
of current technology. A variation of the technique allows for
controlled generation of entangled subsequent pulses, with the
atomic motion serving as intermediate memory of the quantum state.
\end{abstract}


\maketitle

Coherently controlled atom-photon interfaces are the basic
building blocks of implementations of quantum information
processing and secure telecommunication with quantum optical
systems. Experimental efforts towards this goal have reached
several remarkable milestones. For instance, in the optical regime
quantum correlations between atomic gases and light have been
created and explored
\cite{LukinScience03,Kimble03,Giacobino,Polzik04}; entanglement
between a single trapped ion and a single photon has been
demonstrated~\cite{Monroe04}; highly controlled photonic
interaction has been achieved with single atoms flying through
resonators~\cite{An94,MicroCQED,Kuhn02,Kimble-fly}, or with atoms
or ions trapped by an external potential inside an optical
cavity~\cite{Kimble-atomlaser,Kimble-photon,
Guthoehrlein01,Mundt02,Keller04}. The latter experiments have
demonstrated, amongst others, the generation of Rabi oscillations
between the atomic dipole and the cavity
field~\cite{MicroCQED,Mundt02}, laser action at the level of a
single atom~\cite{An94,Kimble-atomlaser}, and the creation of
single photons on demand~\cite{Kuhn02,Kimble-photon,Keller04}. In
particular, in the experiment of Ref.~\cite{Keller04}
single-photon wave packets of adjustable shape and emission rate
were generated. These realizations access novel regimes of
engineering atom-photon interaction and open promising
perspectives for implementing controlled nonlinear dynamics with
quantum optical systems~\cite{Lukin03}. Besides its applications,
this progress touches on interesting fundamental questions, such
as how macroscopic nonlinear phenomena emerge from the dynamics of
single quantum systems.

In this context, we show here that a single cold trapped atom in a
high-finesse resonator can be used for the controlled,
quantum-coherent generation of entangled light pulses, by
exploiting the mechanical effects of atom-photon interaction. The
atom's motional degrees of freedom act as a quantum medium which
is used to establish entanglement between two field modes or to
store and transmit quantum correlations between subsequent light
pulses with variable delay. In all cases, at the end of the
process the quantum medium is perfectly decorrelated from the
electromagnetic field modes.

The specific application that we describe is that after a short
coherent excitation pulse from the laser, the cavity will emit a
pulse of two-mode squeezed, i.e.\ entangled, light. We also
describe a variation of the scheme which allows for controlled
generation of entangled \textit{subsequent} pulses, with the
atomic motion serving as intermediate memory of the quantum state.

Our scheme extends concepts developed for macroscopic
oscillators~\cite{Pirandola03} to a single quantum optical system,
namely to exploit the coupling between internal and external
(motional) degrees of freedom for controlled, coherent creation of
non-classical light. While this coupling is extremely small in
macroscopic systems, in atomic systems it is significant, thus
moving its application for quantum information processing within
experimental reach. Our study is also connected to ideas of
mapping quantum states of atoms onto light inside a
resonator~\cite{Ze-Parkins99,Parkins02}, and to recent experimental and
theoretical studies on quantum correlations in light
scattering~\cite{LukinScience03,Kimble03,Polzik04,Giacobino,Polaritons,Jakob99,NaGutKike}.
The proposal differs fundamentally from
existing methods for generating pulsed squeezing \cite{kumar} or
intense pulses of polarization-entangled photons
\cite{Bouwmeester} which employ nonlinear crystals driven by a
pulsed pump: in our case the microscopic nature of the medium
allows for full coherent control of the light-matter quantum
correlations and of the final quantum state of the generated
light.

\begin{center}
\begin{figure}[tb]
\epsfig{width=0.99\hsize, file=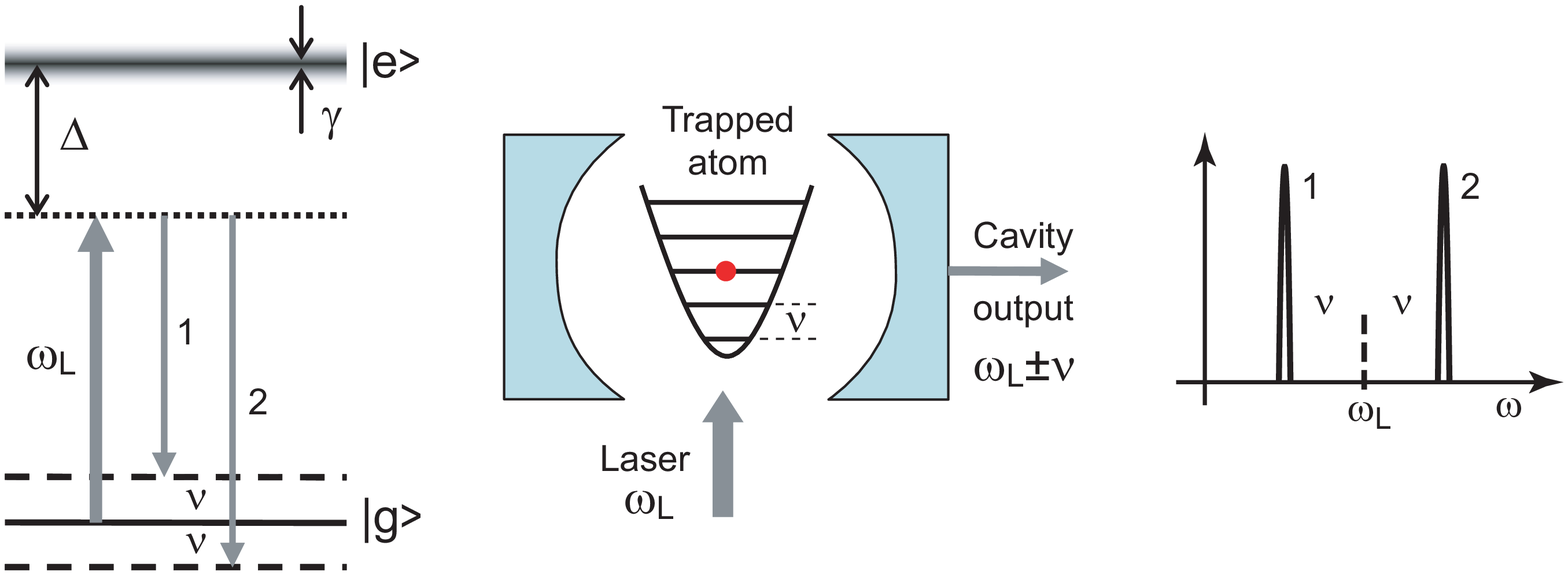} %
\caption{Layout of the system and energy diagram. A single atom
with internal energy levels $|g\rangle$ and $|e\rangle$ is
confined by an external potential inside an optical resonator and
is driven by a laser. The orientation of the vibrational mode
under consideration has a non-zero projection onto the laser
direction. The harmonic motion modulates the laser frequency,
$\omega_L$, and the Stokes and anti-Stokes components at $\omega_L
\pm \nu$ are resonant with two cavity modes, labelled 1 and 2. The
linewidth of $|e\rangle$ is $\gamma$, and $\Delta$ is the detuning
between laser and atom.} \label{Fig:1}
\end{figure}
\end{center}

We first focus on the coherent generation of simultaneous,
entangled light pulses. The physical system is sketched in
Fig.~\ref{Fig:1}. A single trapped atom of mass $M$, for example
an ion in an ion trap \cite{Mundt02,Guthoehrlein01}, is situated
inside a high-finesse optical resonator and is driven by a laser.
The atomic center-of-mass motion is a quantum harmonic oscillator
of frequency $\nu$, described by the operators $b$, $b^{\dagger}$,
which annihilate and create, respectively, a quantum of motional
energy $\hbar\nu$ (a phonon).
The atom's internal transition between states $|g\rangle$ and
$|e\rangle$ has resonance frequency $\omega_0$ and linewidth
$\gamma$ and interacts with a laser at frequency $\omega_L =
\omega_0+\Delta$, where $\Delta$ is the detuning. The transition
couples also with two modes of the resonator at frequencies
$\omega_{1,2} = \omega_L \mp \nu$, which are described by
annihilation and creation operators $a_j$ and $a_j^{\dagger}$
($j=1,2$). These modes can be, e.g., non-degenerate polarization
modes split by $2\nu$. Assuming that the cavity and the laser are
sufficiently far off resonance with the atomic transition,
$|\Delta|\gg\nu,\gamma$, the atom's internal degrees of freedom
can be eliminated from the dynamics, and coherent Raman
transitions are driven between the quantum motion and the cavity
modes. They are described by the effective Hamiltonian $H=H_1+H_2$
with
\begin{eqnarray}
&&H_1={\rm i}\hbar\chi_1 a_1^{\dagger}b^{\dagger}+{\rm
 H.c.}\label{H:1}\\
&&H_2={\rm i}\hbar\chi_2a_2^{\dagger}b+{\rm H.c.} \label{H:2}
\end{eqnarray}
Hamiltonian $H_1$ describes simultaneous creation or annihilation
of a photon in mode 1 and a phonon, while $H_2$ describes the
exchange of excitation quanta between mode 2 and the motion. The
coupling constants for these interactions are
\begin{eqnarray}
&&\chi_1 = \eta
g_1^*\Omega\left(\frac{\cos\theta_L}{\Delta-\nu+{\rm i}\gamma/2}
-\frac{\alpha_1\cos\theta_c}{\Delta+{\rm i}\gamma/2}\right)\\
&&\chi_2 = \eta
g_2^*\Omega\left(\frac{\cos\theta_L}{\Delta+\nu+{\rm i}\gamma/2}
-\frac{\alpha_2\cos\theta_c}{\Delta+{\rm i}\gamma/2}\right)
\end{eqnarray}
where $g_1,g_2$ are the vacuum Rabi frequencies of the cavity
modes, $\Omega$ is the Rabi frequency of the laser, and
$\eta=\sqrt{\hbar k^2/2M\nu}$ is the Lamb-Dicke
parameter~\cite{Stenholm86}, for which we assume $\eta\ll 1$, as
typical for ion traps \cite{footnote_eta}. The geometric terms add
up coherently the mechanical effects of laser and
resonator~\cite{Cirac1993-1995}, whereby $\theta_L$ and $\theta_c$
are the angles between the trap axis and the wave vectors
$\vec{k}_L$ and $\vec{k}_c$ of the laser and the cavity,
respectively ($|\vec{k}_L|\approx |\vec{k}_c|=k$); the scalars
$\alpha_1$, $\alpha_2$ depend on the field gradients of the cavity
modes at the trap center. We will exploit the asymmetry between
$\chi_1$ and $\chi_2$ which originates from the different
detunings, $\Delta \mp \nu$, of the two resonator modes from the
atomic resonance. On a time scale on which the dynamics are
coherent this asymmetry is relevant if $\nu\gg\gamma$.

Let us now assume that a laser pulse of length $T$ interacts with
the atom, whereby $\nu T\gg 1$ in order to ensure spectral
resolution of the two cavity modes. We further assume that during
this interaction incoherent processes such as spontaneous
emission, cavity decay, and dephasing of the center-of-mass motion
can be neglected. The appropriate parameter regime where these
assumptions hold will be discussed below. In this limit the
dynamics are coherent and described by Eqs.~(\ref{H:1})
and~(\ref{H:2}). For $|\chi_2|>|\chi_1|$ the solution of the
corresponding Heisenberg equations are periodic with frequency
$\Theta=\sqrt{|\chi_2|^2-|\chi _1|^2}$~\cite{Pirandola03}. In
particular, after a period $T_{\pi}=\pi/\Theta$ the two cavity
modes exhibit quantum correlations, namely
\begin{eqnarray}
\label{a_1} a_1(T_{\pi})
&=&\frac{|\chi_2|^2+|\chi_1|^2}{\Theta^2}a_1(0)-
2\frac{\chi_1\chi_2}{\Theta^2}a_2^{\dagger}(0)\\
a_2(T_{\pi}) &=& \frac{2\chi_1\chi_2}{\Theta^2}a_1^{\dagger}(0)
-\frac{|\chi_1|^2+|\chi_2|^2}{\Theta^2}a_2(0) \label{b}
\end{eqnarray}
Moreover, the two modes are decoupled from the center-of-mass
oscillator, since $\; b(T_{\pi})=-b(0) \;$. This solution implies
that if at $t=0$ the state of the system is $\rho(0)=\mu\otimes
|0,0 \rangle \langle 0,0|$, where $|0,0\rangle$ is the vacuum
state for both cavity modes and $\mu$ is any state of the
center-of-mass oscillator (for instance a thermal distribution),
then at $t=T_{\pi}$ the state of the system is
$\rho(T_{\pi})=\mu'\otimes |\psi\rangle\langle \psi|$, where
\begin{eqnarray}
|\psi\rangle = \left(\frac{1-r^2}{1+r^2}\right)\sum_{n=0}^{\infty}
\left[-\frac{2r}{1+r^2}e^{i\beta}\right]^n |n,n\rangle \label{nn}
\end{eqnarray}
is a two-mode squeezed state of the two cavity modes, exhibiting
EPR-entanglement~\cite{Reid,Giedke}. In Eq.~(\ref{nn}) the
parameters are $r=\left|\chi_2/\chi_1\right|$,
$\beta=\arg(\chi_1)+\arg(\chi_2)$, and $|n,n\rangle$ describes a
state with $n$ photons in each cavity mode. The average number of
photons per mode is $\langle n\rangle=4r^2/(1-r^2)^2$. Hence, if
the laser pulse has duration $T_{\pi}$, after the interaction the
cavity modes are entangled with one another and decorrelated from
the quantum motion. The motion plays a fundamental role in
establishing the entanglement, nevertheless its initial state does
not affect the efficiency of the process.

Eq.~(\ref{nn}) describes the field inside the cavity at the end of
the laser interaction, provided that $\kappa T_{\pi}\ll 1$, i.e.\
the laser pulse is much shorter than the cavity lifetime
$1/\kappa$. This field is subsequently emitted from the cavity and
contributes to the total cavity output field with a source term
$${\bf E}_s=\sum_{j=1,2}\vec{E}_{0j}a_j(t)+{H.c.}$$
where $a_j(t)\approx a_j(T_{\pi}){\rm e}^{-\kappa t}$ for
$t>0$~\cite{footnote_kappa}. Hence, ${\bf E_s}$ describes a
bichromatic pulse which exhibits two-mode squeezing. The total
output field is ${\bf E}_{\rm out}={\bf E}_s+{\bf E}_f$, where
${\bf E}_f$ is the free field contribution, i.e.\ the external
vacuum modes mixing with ${\bf E_s}$ at the cavity output mirror.
The total output field exhibits the two-mode squeezing of the
source field over a time scale of $1/\kappa$, provided that
$\langle n \rangle \gg 1$.

Analogously, the atomic motion can mediate quantum correlations
between subsequent pulses. A pair of entangled pulses with finite,
variable delay between them can be obtained by coupling the atom
to a single cavity mode at frequency $\omega_c$. In a first step,
a laser pulse tuned to $\omega_c + \nu$ drives the atomic dynamics
according to Hamiltonian $H_1$ (Eq.~(\ref{H:1})) for a chosen
interval of time. At the end of the pulse the motion and the
cavity mode are in a two-mode squeezed state. After a delay
$T_{12}\gg 1/\kappa$, when the first light pulse has left the
cavity, a second laser pulse at frequency $\omega_c - \nu$ drives
the atom according to $H_2$ (Eq.~(\ref{H:2})). With the
appropriate pulse area, the previously created quantum state of
the motion will be transferred to the cavity mode, such that the
motion is decorrelated from the field modes, and the second pulse
leaving the cavity exhibits correlated amplitude fluctuations with
the first one. Hence the motion acts as an intermediate memory for
the quantum state of the first pulse and mediates its entanglement
with the second one, with the memory time limited only by the
decoherence time of the ion motion which can be extremely long.

\begin{center}
\begin{figure}
  \epsfig{width=0.9\hsize, file=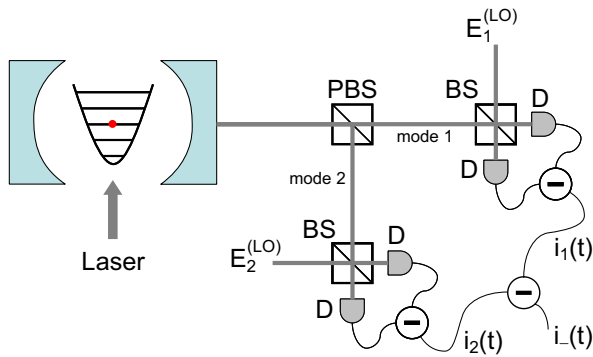}
\caption{Schematic set-up for measuring relative fluctuations
between the two electromagnetic field modes at the cavity output.
PBS stands for polarizing beam splitter, BS for beam splitter, D
for detector. }
  \label{setup}
\end{figure}
\end{center}

A generic experimental set-up to measure the entanglement of
simultaneous pulses is shown in Fig.~\ref{setup}. The two-mode
output field ${\bf E}_{\rm out}$ is split by a polarizing beam
splitter, and the fluctuations of both modes are measured by
balanced homodyne detectors~\cite{Reid}, using local oscillators
$E_{1,2}^{(LO)}$ with phases $\theta_{1,2}$. We define the
corresponding quadratures of the two modes $j=1,2$,
$Q_{j}(\theta_j)=q_{j}+\delta q_{j}$, where $q_{j}=a_{j}{\rm
e}^{{\rm i}\theta_{j}}+ a_{j}^{\dagger}{\rm e}^{-{\rm
i}\theta_{j}}$ are the quadratures of the source fields and
$\delta x_{j}$ denote the corresponding quadratures of the free
fields. The measured currents at the detectors are $i_{j}(t)=
c|E_{j}^{(LO)}|Q_{j}$, where $c$ is a scaling parameter assumed to
be equal for the two modes. Time $t$, starting after the laser
pulse, is considered to be defined on a grid $\delta t$, such that
$\kappa\delta t\ll 1$, i.e.\ fluctuations are recorded on a time
scale much faster than the cavity decay
time~\cite{footnote_deltat}. The correlations are measured through
the difference current $i_-(t)=i_1-i_2=c|E^{(LO)}|\left( Q_{1}-
Q_{2} \right)$, where we have set the two local oscillator
amplitudes equal. The current fluctuations at time $t$ are
$\langle i_-(t)^2 \rangle \propto C(t)$, where
\begin{equation}
C(t)=1-\frac{{\cal R}(t)}{1+{\cal R}(t)}\frac{ \langle q_{1}
q_{2}\rangle}{\langle q_{1}^{2}\rangle + \langle q_{2}^{2}\rangle}
\end{equation}
with ${\cal R}(t)=\kappa\delta t{\rm e}^{-2\kappa t}\left(\langle
q_{1}^{2}\rangle + \langle q_{2}^{2}\rangle\right)$ and
\begin{eqnarray}
&&\langle q_{1}^{2}\rangle=\langle q_{2}^{2}\rangle=
[(|\chi_1|^2+|\chi_2|^2)^2+4|\chi_1\chi_2|^2]/\Theta^4 \qquad \\
&&\langle q_{1} q_{2} \rangle={\rm
Re}\left\{4\chi_1\chi_2(|\chi_1|^2+|\chi_2|^2){\rm e}^{{\rm
i}(\theta_1 + \theta_2)}/\Theta^4\right\}
\end{eqnarray}
and we have assumed that the free field is the coherent vacuum
field. The value $C(t)=1$ corresponds to the shot noise limit for
independent vacuum inputs into the homodyne detectors. The
correlations $\delta(X_1-X_2)^2$ and $\delta(P_1+P_2)^2$ of the
two orthogonal quadratures $X$ and $P$ are obtained by setting
$\theta_1=\theta_2=0$ and $\theta_1=-\theta_2=\pi/2$,
respectively, which leads to identical results for $C(t)$. Thus
$C(t)<1$ is a signature (and in fact, a quantitative measure
\cite{Giedke}) of EPR-type entanglement.

\begin{center}
\begin{figure}
\epsfig{width=0.65\hsize, file=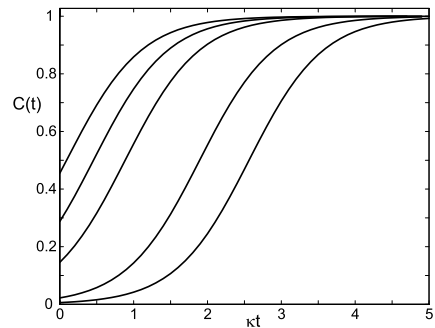} \caption{Signal $C$ as a
function of time, for $\theta_1+\theta_2=0$ and for values of the
parameter $r=1.8,1.5,1.3,1.1,1.05$ (from left to right). A time
resolution of $\delta t = \kappa/10$ has been used. The other
parameters are discussed in the text.} \label{Fig:3}
\end{figure}
\end{center}

Figure~\ref{Fig:3} shows the signal $C(t)$ for different values of
the squeezing parameter~$r$. A reduction below 10\% of the shot
noise level is reached on a time scale of $1/\kappa$ for $r=1.1$.
About 110 photons per mode are created in this case. It should be
noted that for $r$ close to 1, significant two-mode squeezing is
observed over several cavity decay times, before the quantum noise
level $C(t)=1$ is approached when the number of photons remaining
in the cavity reaches the order of one.

We now discuss the parameter regime in which our description
holds. Hamiltonians $H_1$ and $H_2$ 
are valid at first order in the Lamb-Dicke expansion, when
higher-order, off-resonant and inelastic scattering processes are
negligible. Elastic scattering processes have been traced out from
these dynamics, since they do not affect the coherence properties
of the center-of-mass oscillator and of the cavity
modes~\cite{Ozeri2005}. Spontaneous scattering of cavity photons
can be neglected if $\kappa \gg \gamma g_j^2/ \Delta^2$ ($j=1,2$).
The motion is not affected by spontaneous scattering provided that
$\Theta \gg \eta^2 \gamma \Omega^2/\Delta^2$. Cavity decay is
negligible during the interaction if $\Theta \gg \kappa$.
Moreover, Hamiltonians $H_1$, $H_2$ are based on the assumption
that the cavity modes are spectrally resolved, i.e. $\nu\gg
\Theta$. Therefore, one important inequality to be fulfilled is
\begin{equation}
\label{Condition} \nu\gg\Theta\gg\kappa
\end{equation}
Finally, coherence of the center-of-mass oscillator during the
whole time is a prerequisite for the efficiency of this scheme. We
remark that the parameter $r$, which determines the degree of
entanglement, is controlled by the ratio $\nu/\Delta$, provided
that $\nu \gg \gamma$, and by the ratio $g_1/g_2$, i.e.\ the
values of the vacuum Rabi frequencies at the trap
center~\cite{footnote_r}.

The experimental conditions may, for example, be achieved on a
${J}\!=\!1/2 \leftrightarrow {J^{\prime}}\!=\!1/2$ or ${F}\!=0
\leftrightarrow {F^{\prime}}\!=\!1$ atomic transition with the
quantization axis $\vec{B}$ along the cavity axis, and $\vec{B}$,
$\vec{k}_L$, and laser polarization $\vec{E}_L$ mutually
orthogonal. The parameter regime can be accessed by an Indium ion
which is confined by a trap at $\nu=2\pi\times 3$~MHz and whose
intercombination line at $\gamma=2\pi\times 360$~kHz is
laser-driven with Rabi frequency $\Omega=2\pi\times 18$~MHz and
detuning $\Delta=2\pi\times 60$~MHz. The ion couples to two
non-degenerate polarization modes of a resonator of finesse ${\cal
F}=10^6$ and free spectral range $\delta\omega=2\pi\times 1$~GHz.
With the resulting set of parameters
$(g,\kappa,\gamma)/2\pi=(500,1,360)$~kHz, highly entangled pulses,
characterized by $C(t)<0.1$, are observed on a time scale of the
order of 0.1~msec.
%
%
To obtain similar results with subsequent pulses requires a
smaller Lamb-Dicke factor, i.e. a tighter trap. However, only a
single cavity mode is needed, thereby relaxing several demanding
conditions on the properties of the cavity and on the pulse
duration, and thus allowing for a wider range of applications.

To conclude, we have shown that the motion of a single trapped
atom inside an optical resonator can act as a quantum medium which
mediates entanglement on demand between simultaneous or subsequent
radiation pulses. The process is based on the mechanical effect of
light, which in the quantum regime allows for coherently
controlling the interaction and thereby the degree of
entanglement. It can be extended to the microwave regime by
suitably driving atomic microwave transitions in a setup like the
one discussed in~\cite{Wunderlich}. In the future we will study
correlations in the continuous-wave excitation of the ion, in the
perspective of applications for quantum networking, like for
instance discussed in~\cite{CiracKimble,Kraus04}.

The authors gratefully acknowledge discussions with Christoph
Becher, Markus Hennrich, and Scott Parkins. This work was partly
supported by the European Commission (CONQUEST network,
MRTN-CT-2003-505089); G.~M. is supported by the Spanish Ministerio
de Educaci\'on y Ciencia (Ramon-y-Cajal).


\begin{thebibliography}{99}




\bibitem{LukinScience03}
C. H. van der Wal {\it et al.}, Science {\bf 301}, 196 (2003).


\bibitem{Kimble03}
A. Kuzmich {\it et al.}, Nature {\bf 423}, 731 (2003).


\bibitem{Giacobino}
V. Josse {\it et al.}, Phys. Rev. Lett. {\bf 91}, 103601 (2003);
V. Josse {\it et al.}, Phys. Rev. Lett. {\bf 92}, 123601 (2004).


\bibitem{Polzik04}
B. Julsgaard {\it et al.}, Nature {\bf 432}, 482 (2004).


\bibitem{Monroe04}
B. Blinov {\it et al.}, Nature (London) {\bf 428}, 153 (2004).


\bibitem{An94}
K. An {\it et al.}, Phys. Rev. Lett. {\bf 73}, 3375 (1994).


\bibitem{MicroCQED}
J. M. Raimond {\it et al.}, Rev. Mod. Phys. {\bf 73}, 565
(2001); B. T. H. Varcoe {\it et al.}, Nature {\bf 403}, 743
(2000).

\bibitem{Kimble-fly}
C. J. Hood {\it et al.}, Science {\bf 287}, 1457 (2000);
P. W. H. Pinkse {\it et al.}, Nature {\bf 404}, 365 (2000).


\bibitem{Kuhn02}
A. Kuhn {\it et al.}, Phys. Rev. Lett. {\bf 89},
067901 (2002);
T. Legero {\it et al.}, Phys. Rev. Lett. {\bf 93}, 070503 (2004).


\bibitem{Kimble-atomlaser}
J. McKeever {\it et al.}, Nature {\bf 425}, 268 (2003).


\bibitem{Kimble-photon}
J. McKeever {\it et al.}, Science {\bf 303}, 1992 (2004).


\bibitem{Guthoehrlein01}
G. R. Guth\"ohrlein {\it et al.}, Nature {\bf 414}, 49 (2001).


\bibitem{Mundt02}
A. B. Mundt {\it et al.}, Phys. Rev. Lett. {\bf 89}, 103001
(2002).


\bibitem{Keller04}
M. Keller {\it et al.}, Nature {\bf 431}, 1075 ( 2004).

\bibitem{Lukin03}
M. D. Lukin, Rev. Mod. Phys. {\bf 75}, 457 (2003).







\bibitem{Pirandola03}
S. Mancini {\it et al.}, Phys. Rev. Lett. {\bf 90}, 137901 (2003); S.~Pirandola
{\it et al.}, Phys. Rev. A {\bf 68}, 06 2317 (2003).


\bibitem{Ze-Parkins99}
H. Zeng and F. Lin, Phys. Rev. A {\bf 50}, R3589 (1994); A. S. Parkins and H.
J. Kimble, J. Opt. B: Quantum Semiclass. Opt. {\bf 1}, 496 (1999).

\bibitem{Parkins02}
Peng and A.S.\ Parkins, Phys.\ Rev.\ A {\bf 65}, 062323 (2002).


\bibitem{Polaritons}
J. Ph. Karr {\it et al.}, Phys. Rev. A {\bf 69}, 063804
(2004).

\bibitem{Jakob99}
M.~Jakob, J.~Bergou, Phys.\ Rev.\ A {\bf 60}, 4179 (1999).

\bibitem{NaGutKike}
R. Guzm\'{a}n \textit{et al.}, quant-ph/0407218.


\bibitem{kumar}
M. E. Anderson {\it et al.}, J. Opt. Soc. Am. B \textbf{14} 3180 (1997);
G. S. Kanter, {\it et al.}, Opt. Expr. \textbf{10} 177 (2002).


\bibitem{Bouwmeester}
C. Simon and D. Bouwmeester, Phys. Rev. Lett. \textbf{91}, 053601 (2003).



\bibitem{Stenholm86}
S. Stenholm, Rev. Mod. Phys. {\bf 58}, 699 (1986).


\bibitem{footnote_eta}
The condition $\eta \ll 1$ can be significantly relaxed because in
the configuration we describe, only the first-order motional
sideband couples to the cavity.


\bibitem{Cirac1993-1995}
J. I. Cirac \textit{et al.}, Phys. Rev. A {\bf 48}, 2169 (1993);
J. I. Cirac {\it et al.}, Phys. Rev. A {\bf 51}, 1650
(1995).


\bibitem{Reid}
M. D. Reid, Phys. Rev. A {\bf 40}, 913 (1989).


\bibitem{Giedke}
G. Giedke {\it et al.}, Phys. Rev. Lett. {\bf 91}, 107901 (2003).




\bibitem{footnote_kappa}
We assume that both modes decay at rate $\kappa$ and set the
cavity output mirror at the spatial origin of the axis.

\bibitem{footnote_deltat}
The inverse of $\delta t$ corresponds to the spectral bandwidth
over which vacuum fluctuations contribute to the signal.

\bibitem{Ozeri2005}
R. Ozeri \textit{et al.}, Phys.\ Rev.\ Lett.\ {\bf 95},
030403 (2005).

\bibitem{footnote_r}
If $\nu<\gamma$, very large values of $\Delta$ are required. In
this case, $r$ can still be controlled by the ratio $g_1/g_2$.


\bibitem{Wunderlich}
F.~Mintert and Ch.~Wunderlich, Phys.\ Rev.\ Lett.\ {\bf 87},
257904 (2001).

\bibitem{CiracKimble}
J. I. Cirac {\it et al.}, Phys. Rev. Lett. {\bf 78}, 3221 (1997).


\bibitem{Kraus04}
B. Kraus and J. I. Cirac, Phys. Rev. Lett. {\bf 92}, 013602
(2004).


\end{thebibliography}
\end{document}